\documentclass[acmsmall]{acmart}

\newcommand{\case}[2]{{%
\textit{``#2''}}}

\newcommand{\caseTable}[2]{{%
\textit{#2}}}

\newcommand{\tosdr}[1]{{%
ToS;DR}}

\newcommand{\qOne}[1]{What is your age?}
\newcommand{\qTwo}[1]{What is your occupation?}
\newcommand{\qThree}[1]{What is your Nationality?}
\newcommand{\qFour}[1]{Q1. How motivated are you to read privacy policies before using an online service to make sure your information is safe?
(1 means `Not motivated at all' and 10 means `Extremely motivated')}
\newcommand{\qFive}[1]{Q2. How likely are you to invest time in understanding privacy policies to avoid potential privacy issues?
(1 means ‘Very unlikely’ and 10 means ‘Very likely’)}
\newcommand{\qSix}[1]{Q3. Assume you are reviewing a privacy policy, do you focus more on specific sections or a general overview?
(1 means ‘Focus much more on specific sections’ and 10 means ‘Focus much more on general overview’)}
\newcommand{\qSeven}[1]{Q4. How confident are you in your ability to manage privacy settings on social media platforms?
(1 means `Not confident at all’ and 10 means ‘Extremely confident’)}
\newcommand{\qEight}[1]{Q5. How comfortable are you with using technologies that protect your privacy, such as virtual private networks (VPNs), or inspect your privacy hygiene, such as Ghostery?
(1 means ‘Not comfortable at all’ and 10 means ‘Very comfortable’)}
\newcommand{\qNine}[1]{Q6. How important is it for you to feel in control of your personal information online?
(1 means ‘Not important at all’ and 10 means ‘Extremely important’)}
\newcommand{\qTen}[1]{Q7. How proactive are you in updating your privacy settings regularly?
(1 means ‘Never proactive’ and 10 means ‘Always proactive’)}
\newcommand{\qEleven}[1]{Q8. How comfortable are you with sharing personal information with online services you use regularly?
(1 means ‘Not comfortable at all’ and 10 means ‘Very comfortable’)}
\newcommand{\qTwelve}[1]{Q9. Despite the risks, how often do you accept terms and conditions before reading them?
(1 means ‘Never’ and 10 means ‘Always’)}
\newcommand{\qThirteen}[1]{Q10. How likely are you to take action (e.g., change settings, stop using a service) if you find a privacy policy concerning?
(1 means ‘Very unlikely’ and 10 means ‘Very likely’)}
\newcommand{\qFourteen}[1]{Q11. Suppose educational resources or tools were available to help you better understand privacy policies, how much would you be interested to learn more about such resources?
(1 means ‘Not interested at all’ and 10 means ‘Extremely interested’)}
\newcommand{\qFifteen}[1]{Have you ever had any of the following negative experiences due to not understanding or not reading a privacy policy?
(E.g., Account getting hacked, Data used elsewhere, Account discontinued, no refund, etc.)\\
Account Hacked\\
Data Breach\\
Received recommendations without opting-in\\
Received personalized ads without opting-in\\
Failed to obtain a refund\\
Observed unexpected charges\\
Account suspended/terminated\\
Misunderstood your rights or responsibilities\\
Data Misused in some other way\\
Other (Please specify)}

\newcommand{\revised}[1]{{%
#1}}

\newcommand{\REDACT}[1]{$\Box REDACTED \Box$} 

\newcommand{\redactCollege}[1]{[a U.S. University]}  

\newcommand{\quotateInset}[4]{%
\vspace{-3pt}%
\begin{quote}%
     \leftskip-10pt
     \rightskip-15pt
\textbf{#1}: \emph{``#4''}\end{quote}%
\vspace{3pt}}

\newcounter{boldifyCounter}
\newcounter{fixmeSectionCounter}
\newcounter{fixmeTotalCounter}
\makeatletter
\@addtoreset{fixmeSectionCounter}{section}
\@addtoreset{fixmeSectionCounter}{subsection}
\@addtoreset{boldifyCounter}{section}
\@addtoreset{boldifyCounter}{subsection}
\makeatother

\newcommand{\boldify}[1]{}
\ifdefined\boldifyON
	\renewcommand{\boldify}[1]{
        \par\noindent
		\stepcounter{boldifyCounter}
		\textbf{{\color{green}**}
		~\arabic{section}.\arabic{subsection}.\arabic{boldifyCounter}
		: #1} 
	}
\fi 

\newcommand{\reportOnFIXME}{%
    \newcount\iterCounter
    \iterCounter=1
    \newcount\endCounter
    \endCounter=\totvalue{fixmeTotalCounter}
    \advance \endCounter +1
    There are 
    {\color{red}\total{fixmeTotalCounter}} 
    FIX\_ME\\
    links:
    \loop
        \hyperlink{fixTag\the\iterCounter}{\#\the\iterCounter}
        \advance \iterCounter +1
    \ifnum \iterCounter < \endCounter
    \repeat
}

\newcommand{\FIXME}[1]{} 
\ifdefined\fixmeON
	\renewcommand{\FIXME}[1]{\par\noindent
		\stepcounter{fixmeSectionCounter}\stepcounter{fixmeTotalCounter}
		{\color{red}\fbox{\color{black}
			\parbox{.965\linewidth}{
				\textbf{\hypertarget{fixTag\thefixmeTotalCounter}{FIXME}	\arabic{section}.\arabic{subsection}.
        		\arabic{fixmeSectionCounter} (\color{red}
        		\#\arabic{fixmeTotalCounter}):} #1}}
        }
	}
\fi 

\newcommand{\FIXED}[1]{}
\ifdefined\fixmeON
	\renewcommand{\FIXED}[1]{\par\noindent%
		{\color{black}\fbox{\color{black}%
			\parbox{.99\columnwidth}{%
				\color{blue}#1}}%
        }
	}
\fi 

\newcommand{\draftStatus}[1]{}
\ifdefined\draftStatusON
	\renewcommand{\draftStatus}[1]{
        \hfill **#1
	}
\fi 

\usepackage{calc, enumitem} 
\usepackage{totcount}\regtotcounter{fixmeTotalCounter}
\usepackage{soul}
\usepackage{adjustbox}
\usepackage{multirow}
\usepackage{enumitem}
\usepackage{colortbl}
\usepackage{hyperref}
\usepackage{longtable}
\usepackage{svg}
\usepackage{datetime} 

\setcopyright{none}
\copyrightyear{2025}
\acmYear{2025}
\acmDOI{99.9999/9999999.9999999}

\begin{document}

\title{
Signed, Sealed,... Confused: Exploring the Understandability and Severity of Policy Documents
}

\author{Shikha Soneji}
\email{sxs7000@psu.edu}
\affiliation{%
  \institution{%
  Penn State University}
  \streetaddress{Westgate Building}
  \city{University Park}
  \state{PA}
  \country{USA}
  \postcode{16802}
}

\author{Sourav Panda}
\email{sbp5911@psu.edu}
\affiliation{%
  \institution{%
  Penn State University}
  \streetaddress{Westgate Building}
  \city{University Park}
  \state{PA}
  \country{USA}
  \postcode{16802}
}

\author{Sameer Neve}
\email{FIXME}
\affiliation{%
  \institution{Independent Researcher}
  \city{Wayne}
  \state{NJ}
  \country{USA}
  \postcode{07470}
}

\author{Jonathan Dodge}
\email{jxd6067@psu.edu}
\affiliation{%
  \institution{%
  Penn State University}
  \streetaddress{Westgate Building}
  \city{University Park}
  \state{PA}
  \country{USA}
  \postcode{16802}
}


\renewcommand{\shortauthors}{S.\ Soneji, S.\ Panda, S.\ Neve, and J.\ Dodge}

\begin{abstract}
In general, Terms of Service (ToS) and other policy documents are verbose and full of legal jargon, which poses challenges for users to understand.
To improve user accessibility and transparency, the ``Terms of Service; Didn't Read'' (ToS;DR) project condenses intricate legal terminology into summaries and overall grades for the website's policy documents.
Nevertheless, uncertainties remain about whether users could truly grasp the implications of simplified presentations.
We conducted an online survey to assess the perceived understandability and severity of randomly chosen cases from the ToS;DR taxonomy.
Preliminary results indicate that, although most users report understanding the cases, they find a bias towards service providers in about two-thirds of the cases.
The findings of our study emphasize the necessity of prioritizing user-centric policy formulation.
This study has the potential to reveal the extent of information imbalance in digital services and promote more well-informed user consent.
\end{abstract}


%

\maketitle

\section{Introduction
}\label{secIntroduction}

By failing to read privacy policies, users risk a range of potential hazards.
For instance, people may unintentionally authorize data sharing with third parties, which might result in unwanted marketing or worse.
Milne et al.~\cite{milne2004consumers} revealed that a substantial number of users lack awareness of the collection and utilization of their personal information.
Insufficient knowledge about this matter can lead to users relinquishing authority over their personal data, resulting in a variety of consequences.
One example is \textit{Feldman v.\ Google, Inc.}~\cite{2007feldman}, which centers on Google AdWords and unexpected costs incurred by the plaintiff due to click fraud.
This case highlights the difficulties consumers have when they fail to carefully read (and/or comprehend) such agreements.
Another example is \textit{Fteja v.\ Facebook, Inc.}~\cite{2012fteja}, which refers to an allegedly unjustified account deactivation.
The case highlights what happens when individuals accept agreements from digital platforms without fully understanding them, especially their rights and access to social media sites.

\boldify{Interventions for this problem are plentiful, but generally insufficient}

There have been continuous efforts to streamline privacy policies, with varied levels of success.
One solution is \textit{layered policies}, which offer a concise overview alongside links to more comprehensive information.
However, prior work~\cite{mcdonald2009comparative} showed that although layered policies have benefits, users must still exert effort to acquire and comprehend the complete information.
An alternative approach entails symbolizing essential privacy standards with icons~\cite{kelley2010standardizing}.
These visual aids facilitate users' rapid comprehension of the fundamental aspects of a policy; however, they may oversimplify legal terminology.
The practice of simplifying legal documents on websites such as Pinterest~\cite{pinterest},
500px~\cite{500px}, and 
Prolific~\cite{prolific}
stands as proof of the advancement of user-centric design on digital platforms.

\boldify{One major complication is this paradox between reported and actual behavior, in terms of privacy preferences}

The discrepancy between the user's interest in knowing about data collection and their behavior is called the Privacy Paradox~\cite{barnes2006privacy}.
According to prior work~\cite{acquisti2015privacy}, users express concerns about their privacy, but their behavior often contradicts these concerns.
Those authors attribute the contradiction to the complicated and lengthy nature of privacy policies and highlight the necessity for the development of privacy policy presentations that are more accessible to 
\revised{%
people without specific privacy training.
} 
This paradox is apparent in other studies (e.g., \cite{norberg2007privacy}) which demonstrate that despite users asserting the importance of their privacy, they frequently reveal sensitive information in exchange for small incentives or conveniences.
The inconsistency between statements and behavior is an obstacle to resolving privacy concerns in the digital era.

\boldify{And so, we wind up in a mess, with human-centered approaches being a possible way forward}

Unfortunately, existing solutions frequently fail to understand or address the underlying causes of the Privacy Paradox.
Regulations still largely depend on legalese and are not accessible to
\revised{%
individuals with limited knowledge about privacy.
} 
Prior work~\cite{reidenberg2015disagreeable} found that even simpler policies are difficult to comprehend with low legal literacy.
This highlights the necessity for implementing more human-centered methods in the development of privacy policies.
\revised{%
One approach driven by human expertise is Terms of Service; Didn't Read (\tosdr{})~\cite{tosdr}.
Over time, \tosdr{} moderators have curated a taxonomy of privacy concepts and the community has submitted fragments of privacy policies that exhibit a particular concept, which the moderators can then approve or deny.
This amounts to a data labeling process, and though the labels are sparse, their sum total is a valuable indicator of how the policy manages the power balance between user and service provider.
The \tosdr{} team grades each website from A (best) to E (worst) based on their policy documents, such as privacy policy, terms of service, data policy, cookie policy, etc.
The detection and scoring of the individual cases found in a document gives its overall grade.
We chose this particular taxonomy since it is expert-derived and has maintained a long-lived crowdsourcing effort, suggesting a mix of high quality and accessibility to broad audiences.

\boldify{set the scope of what we are doing just before RQs}

We seek to understand how people perceive the taxonomy from \tosdr{} in two main respects:
The first metric is how well the participants report being able to understand the concept underlying individual taxa, which we call \textit{understandability}.
The second metric is the extent to which the concept favors a particular party, either the service provider or user, which we call \textit{severity}.
To collect data on these metrics, 
} 
we conducted an online survey to answer the following research questions:
\begin{enumerate}
\item[\textbf{RQ1}] What does the \textit{understandability} data reveal about participants and privacy concepts?
\item[\textbf{RQ2}] What does the \textit{severity} data reveal about participants and privacy concepts?
\revised{%
\item[\textbf{RQ3}] How do participants' severity and understandability ratings for privacy concepts relate to each other?
\item[\textbf{RQ4}] How did participants' demographic traits and attitudes affect their understandability and severity ratings of privacy concepts?
} 
\end{enumerate}

\boldify{Then, the obligatory contribs list, since often need the structure to know what we did}

By answering and discussing these RQs, we make the following contributions:
(1)~Knowledge about the usefulness of case labels as explanatory ingredients, based on understandability data.
\revised{%
(2)~Knowledge about perceptions of severity of the concepts the case labels represent.
While the information listed in these two contributions is useful in its own right, the severity data can also provide a user-perception-based scoring system.
(3)~With the previous three contributions, we offer a basis for future RQs and hypotheses, particularly in targeting educational interventions.
(4)~Alternative candidate text for the cases with low understandability, based on participants rewritten the cases.
} 

\section{Background and Related work
\draftStatus{barring audits, R+R DONE}
}

\revised{%
Understanding privacy policies is crucial for empowering users to make informed decisions about their personal data.
This section explores the challenges users face with privacy policies, the role of social media and data collection practices in shaping attitudes, and the psychological and behavioral factors that influence engagement.
It also highlights innovative approaches that aim to enhance user comprehension and trust, such as annotated policies and visual aids.

\boldify{How comfortable are folks with data collection/sales/etc?}

Websites profit from people's trust by selling, sharing, or analyzing personal information~\cite{torbert2021because}.
Goldfarb and Que~\cite{goldfarb2023economics} explore the significant economic value of digital privacy, demonstrating how consumer privacy preferences shape business strategies and market dynamics.
They also examine the impact of privacy regulations on competition and innovation, highlighting the role of policy in fostering privacy-preserving technologies and balancing consumer protection with economic growth.
Melicher et al.~\cite{melicher2016preferences} discovered that slightly over half of the participants (51\%) felt that information about their general searching activities significantly impacted their feelings about tracking.
This finding underscores the importance of addressing user concerns in privacy policy design to enhance trust and compliance.
} 
Another study has found that websites' failure to protect consumers' personal information has harmed their trust in the company~\cite{pilton2021evaluating}.
Americans have varied attitudes about companies using their personal information to build new products: 50\% of people are at least somewhat comfortable, while 49\% are extremely uncomfortable~\cite{lippi2019claudette}.
The general public is more comfortable with companies using their personal information for some purposes than for others.
As an example, some people are extremely or somewhat comfortable with businesses using their personal information to help them improve their fraud protection systems~\cite{auxier2019americans}.

\subsection{The role of social media apps}

\boldify{Social Media is popular, but a big data vacuuming system}

Facebook, Instagram, Twitter, and TikTok are now common tools for communication, information retrieval, and entertainment, 
\revised{%
which exerts a strong pull on attitudes about data collection.
} 
Statista~\cite{statista} noted that as of 2022, the global user base of social media platforms exceeded 4.26 billion individuals, emphasizing their widespread presence.
The fundamental impetus for using social media is the desire for social interaction.
According to Ellison et al.~\cite{ellison2007benefits}, social networking sites (SNS) offer users the chance to preserve current relationships, establish new connections, and develop social capital.
Hermida et al.~\cite{hermida2012share} observed that Twitter has a substantial impact on spreading news and enabling immediate conversations about current events.
Social media's ability to quickly and widely disseminate information is the core added value, but it also raises questions about the authenticity and dependability of the shared content.

\boldify{Ad targeting is a big factor in the social media profit model}

The monetization of social media applications also has a crucial impact on users' experience.
These platforms specifically target consumers with customized advertisements based on their online behavior and preferences.
Social media marketing has become an essential element of corporate strategies, enabling organizations to directly interact with consumers and cultivate brand loyalty~\cite{kaplan2010users}.
Although targeted ads can improve the user's experience by offering appropriate material, they also give rise to privacy concerns about the data collection that informs targeting.
Golbeck and Mauriello~\cite{golbeck2016} conducted a study on the privacy concerns of Facebook users to gauge how informed they are about the use of their data.
The study revealed that initially, users were not aware of how platforms use their data, and once that information surfaced, their concern about data privacy increased---with differences in the educational value of their treatments:
reading the document (worst results), interacting with an app, and a film with horror stylings (best results).

\subsection{User's knowledge and attitudes about data collection practices}

\boldify{What factors seem to make people more or less willing to opt in?}

\revised{%
According to Earp et al.~\cite{earp2005examining}, people are more likely to read and understand privacy regulations when they perceive a significant risk to their personal data.
This conclusion flows from their examination of user attitudes toward privacy policies, which revealed that individuals' involvement with these policies increases when they feel that their data is under threat.
Additionally, Earp et al.\ found that users' trust in the organizations responsible for the policies plays a critical role, as higher trust often correlates with lower scrutiny of the policy details.
} 
Bansal et al.~\cite{bansal2010impact} emphasize that implementing transparent and unambiguous privacy practices can establish confidence.
In contrast, confidence decreases when people think that the organization lacks openness or has a poor track record.
Those authors demonstrate that perceived reliability and trustworthiness play a crucial role in shaping user attitudes toward data collection, ultimately affecting their inclination to share personal information~\cite{bansal2010impact}.
Occasionally, trust can override privacy concerns, resulting in a misleading perception of safety.
For example, researchers examined the Facebook-Cambridge Analytica scandal, emphasizing its impact on user data privacy and the critical need for enhanced privacy protection measures~\cite{isaak2018user}.

A multitude of other hacks and data breaches demonstrate that even well-regarded institutions can mishandle or inadequately safeguard user data.
We will provide a few famous and large-scale examples
In 2017, Equifax, a prominent credit reporting agency, suffered a security breach that compromised the confidentiality of sensitive personal data, such as Social Security numbers, addresses, and birth dates, belonging to more than 147 million individuals~\cite{bernard2017equifax}.
In the same year, Sony had a massive security breach on its PlayStation Network, resulting in the exposure of personal information, including credit card data, of 77 million customers~\cite{wilton2017sony}. 

\boldify{So if a barrier is actually reading/knowledge, why don't they do that?}

Many users are frequently oblivious to the full scope of data collection conducted by different platforms and services.
Milne et al.~\cite{milne2004consumers} indicates that a substantial proportion of users lack awareness or comprehension of data-gathering processes.
Prior work indicates that a substantial number of users fail to read privacy policies because they find them too complicated and lengthy~\cite{acquisti2015privacy}.
An empirical analysis found that only a few participants read all or almost all of the terms' main reason for not reading being, boring and tedious to read~\cite{maronick2014consumers}.
According to a study from Smith et al.~\cite{smith1996information}, individuals who know the methods of collecting data typically experience a feeling of intrusion and a lack of control over their personal information.
Further, users may provide personal information impulsively, without considering the long-term privacy risks.
Taddicken~\cite{taddicken2014privacy} discovered that the convenience and incentives provided by these platforms frequently surpass privacy worries, resulting in a greater probability of users revealing personal information.
Heuristic cues such as mental shortcuts in the online world may subconsciously lead users to give personal information despite privacy concerns~\cite{sundar2013unlocking}.
Last, users frequently imitate the actions of their peers, particularly in social networking settings, where the sharing of personal information is widespread.
According to Debatin et al.~\cite{debatin2009facebook}, the need for social recognition and a sense of belonging can motivate users to participate in activities that compromise their privacy.

\boldify{Attitudes also seem to vary with data type}

The emotional reaction to data gathering is also affected by the perceived sensitivity of the data under consideration.
Users exhibit heightened worry about gathering personal and sensitive data, including financial information, medical records, and location data.
Based on a survey from Pew Research Center~\cite{pew2019privacy}, most users felt uncomfortable about the collection of their sensitive information.
Many of them had concerns about the possibility of data being misused or accessed without authorization.
To counteract this discomfort, incorporating privacy by design principles throughout the construction of digital services can effectively reduce privacy risks and ensure that user behaviors are in line with their stated privacy concerns~\cite{cavoukian2009privacy}.
The ToS has major legal implications, regardless of whether users understand them, as another court case demonstrates~\cite{2012davis}.
This lawsuit pertained to accusations against HSBC and Best Buy for deceiving California customers by failing to adequately disclose an annual fee for credit cards.

\subsection{Users’ Opinions on Annotated Policies}

Offering users transparent and succinct information regarding the gathering and utilization of data might enable them to make well-informed choices.
The main objective of these annotations is to improve user understanding and provide informed consent by simplifying complex legal terminology into more accessible portions.
Research suggests that the inclusion of explanatory privacy rules greatly enhances consumers' comprehension and retention of information regarding data practices.
For example, Kelley et al.~\cite{kelley2010standardizing} showed that individuals who interacted with simplified privacy notifications better remembered the privacy practices of the websites they visited.
Further, people broadly misunderstand technical terms in privacy policies, reducing transparency and user comfort, highlighting the need for simpler language to enhance policy acceptance~\cite{Tang2021defining}.
To address these concerns, other research has found annotated privacy rules can enhance comprehension of important information~\cite{reidenberg2015disagreeable}.
However, simplification can have both positive and negative consequences, as it makes policy more accessible, but may also result in the omission of important elements.
For example, Bravo et al.~\cite{bravo2010bridging} discovered that some users held a misguided perception of safety, mistakenly thinking they have a complete understanding of a policy, while they know a portion (the annotated sections).

The level of trust that users have in annotated privacy policies likewise varies considerably.
According to Bansal et al.~\cite{bansal2010impact}, annotations have the potential to enhance trust when users perceive the content as sincere and transparent attempts to convey privacy standards.
On the other hand, if users have suspicions that annotations are deliberately creating a positive image by cherry-picking content, confidence can diminish.
Other researchers showed that the presence of transparency and perceived honesty notably influence consumers' acceptance and trust towards annotated privacy regulations~\cite{schaub2015design}.
\revised{%
One such transparency approach is from Shvartzshnaider et al.~\cite{shvartzshnaider2020beyond}, who proposed a framework for analyzing privacy policies using syntactic and semantic role labeling to extract privacy parameters such as sender, recipient, subject, and information type, grounded in the Contextual Integrity framework.
Their approach, tested on 36 real-world privacy policies, achieved an F1 score of 84\%, demonstrating the effectiveness of incorporating domain-specific knowledge into NLP models for privacy policy analysis.
Some efforts at measuring or increasing user understanding of privacy concepts specifically consider children, for example supporting improving children's comprehension of digital privacy via the same framework~\cite{kumar2020strengthening}.
} 

Prior work observed a user preference for visual aids, such as icons and color-coded sections since they facilitated a rapid understanding of essential information~\cite{balebako2022nudging}.
However, depending on images without adequate explaining text can result in shallow comprehension, so it is important to accommodate various learning styles.
As an example, Milne et al.~\cite{milne2004consumers} emphasized that younger individuals exhibit greater receptiveness towards annotations and perceive them as more beneficial, in contrast to older individuals who often prefer conventional text styles.

\section{Methodology
\draftStatus{few fixes remain, revision highlights are in}
}

\boldify{at a high level, what did we do and who did it?}

In the third quarter of 2023, we crawled and scraped data via the \tosdr{} streaming API and BeautifulSoup~\cite{beautifulSoup}.
This data set contains numerous attributes, such as \textit{Description}, \textit{Case}, \textit{DocType}, \textit{Title}, \textit{Status}, \textit{Comments}, and \textit{Authors}.
\revised{%
While the scrape provides labels for fragments of privacy policies, this study will focus on just the \emph{Case} attribute: text that the \tosdr{} team has developed for a privacy concept, about a sentence in length.
This team has developed 245 such cases, which have more detail in a longer paragraph describing the case and offer a ``summary'' of the main ideas in the document.
} 
However, our survey will only focus on 243 of these cases because there were two cases that the \tosdr{} team had not confirmed are actually present in any documents since they rejected all submitted identifications.

In order for people to learn about their ability to understand privacy-related concepts, as well as their impression of how negative or positive that concept was to the affected parties, we conducted a crowd worker survey%
\footnote{The survey is still available at \REDACT{\url{https://severityandunderstandability.ist.psu.edu/}}.}.
To that end, upon obtaining IRB approval, we recruited 519 individuals from Prolific, with inclusion/exclusion criteria that they must speak English fluently and reside in the USA.
The survey contained one attention check question, which would cause rejected work.
\revised{%
Survey responses took FIXME minutes and we paid \$FIXME for accepted work, for an average pay rate of \$FIXME/hr.
} 
Out of the 519 participants, we rejected work from 19 of them and could not recover data from one more, leaving 499 participants.
\revised{%
We chose this sample size based on budget constraints, with a focus on understanding groups of people, meaning we had no intent to compute comparative statistics between cases.
The average time taken by the participants to fill this survey was about 5 minutes and the payment was \$3.} 

\boldify{What core questions did we ask?}

The main task of the survey was to evaluate five random cases.
For each case, we presented the case label and then asked several questions about it.
We first asked about understandability: \textit{``On a scale of 1 to 10, how well do you understand this case? (1 means `Not understanding at all' and 10 means `Understanding this case completely')''}.
When asking about severity, we split the question into two pieces.
The first piece asked: \textit{``Which party does this case tend to favor?''} and participants responded by checking one of three mutually exclusive boxes: \texttt{User}, \texttt{Service Provider}, or \texttt{Neither}.
The second piece then asked: \textit{``How severe is the favoritism that party gains from this case?
(1 means `Not severe at all' and 10 means `Very severe')''}.
\revised{%
We chose this two-step question format to help clarify the participants' thinking by letting them subdivide the problem, akin to determining a vector's direction and then its magnitude.
} 
For the last case, participants saw an additional question that asked them to rewrite the case based on both the case label and its description (\textit{Rewrite the last case (Case \#) in your own words.}).
The minimum requirement for text entry was 21 characters, which helps ensure that participants make a meaningful response.

\boldify{How did the survey flow and what are its constituent parts?}

After obtaining informed consent, the first page collects demographic information from the participant.
The second page gives instructions for the task alongside a single dummy case to clarify the task.
In particular, they answered our questions by checking boxes, repositioning sliders from 0 into the range 1--10, and rewriting one case.
Participants were not able to advance to the next page if any sliders remained at 0 since we consider default values a non-response.
Last comes the page containing the main task (i.e., the five random cases and the aforementioned questions for each case).
The main task page contains one attention check question, which was \textit{``Set the slider to position 7 (This is a special question not related to any case.)''}.

\subsection{Demographic and Attitude Information}

\boldify{Now we describe what demographic information we collect and where we got the questions}

In order to learn about what human traits might moderate responses, we asked about fundamental information such as age, occupation, student status, and gender\footnote{Though we provided multiple gender options in accordance with best practices~\cite{scheuerman2020gender}, nearly all participants identified as male/female, with a few preferring not to state.
} 

We also posed 11 questions to discover their level of awareness of privacy policies, listed later in Table~\ref{table: demoStats}.
We based several questions on language and categories from the US Census~\cite{UScensus}.
Furthermore, we adapted 11 questions adapted from GenderMag~\cite{burnett2016gendermag}, which specifically identifies five facets of users' problem-solving style: motivation, information processing style, learning style, computer self-efficacy, and risk aversion.

\subsection{Analysis}
\label{secAnalysis}

\boldify{how did we prepare the data?}
In order to collect the information on understandability, we asked participants their perceived understanding of each case presented to them, collecting the response on a Likert scale [1, 10].
We then calculated descriptive statistics to assess the understandability with respect to participants, as well as the cases.
After collecting the data, we applied a simple transformation to combine the two questions about severity.
We interpret the response to the second question (which party the case favors) as giving the \emph{sign}, while the Likert response on [1, 10] range gives the \textit{magnitude}.
Therefore, if a participant stated a case favored the \texttt{User}, we would use their severity response directly as a positive value.
However, if they said a case favored the \texttt{Service Provider}, we would negate their severity response.
Finally, if they said a favored \texttt{Neither}, then we would multiply their severity response by 0.
Ultimately, this creates a severity range from [-10, 10], with 10 indicating the case fully favors the User, and -10 fully in favor of the Service Provider.
However, at times we considered the absolute value to assess the statistics with respect to individual cases, each of which are marked clearly in the results.
We analyzed severity similar to understandability, with respect to participants and cases.
We also correlated understandability and severity to assess their impact on each other.

\boldify{And then what kind of analysis did we do?}

Much of our analysis relies on descriptive statistics, including mean, median, standard deviation, count of extremities (values of 1 and 10), and a direct comparison between the scores of understandability and severity and their corresponding means.
Since we used different lenses (per participant, per case, etc.), we extracted data from the raw data in different ways to support each perspective, resulting in different sets of data specific to the analysis variable.
\revised{%
Our last quantitative analysis was to perform hypothesis testing about the influence of demographic traits on participant responses.
Specifically, we checked normality with the Kolmogorov-Smirnov test and equivariance with Levene's test.
Fortunately, none of the distributions violated assumptions and we were able to proceed with parametric statistics, such as two-sample t-test and one-way ANOVA.
} 


\section{Results RQ1 - What does the \textit{understandability} data reveal about participants and privacy concepts?
\draftStatus{one fix remains, revision highlights are in}
}

\boldify{Plot the overall course of the argumentation (here the comma-separated list is roughly the section heading list)}

\revised{%
In this section we report results based on the understandability data, interpreting the grand mean, case mean, case variances, and qualitative results.

\boldify{Overall, the ratings were high, suggesting understandability.
Also, the results were not outlier-driven, because we see a majority of SOMETHING being high scores.}

Overall, we find the descriptive statistics of understandability scores aggregated across all cases and participants indicate that \textbf{\textit{participants reported that the concepts represented by the overall case label set were understandable}}, with moderate variation (Mean=7.42, Median=8, SD=2.71).
Further, a large portion (>72\%) of responses rated the cases as highly understandable with a rating of 7 or more, which indicates that \textbf{most participants found the content to be understandable}.
These findings align with prior research indicating that simplified and annotated versions of privacy policies enhance user comprehension~\cite{kelley2010standardizing, reidenberg2015disagreeable}.  

\boldify{HOWEVER, we do see some low scores, which means some mixtures of case+individuals did not work}

However, despite the high overall ratings, some cases remained unclear for some individuals.
About 5\% of cases received at least one response of 1, indicating very low understandability, while 31.66\% cases got the highest score of 10 from at least one participant.
Additionally, 83.93\% of the responses rated the understandability $\geq5$, indicating that the vast majority of participants found the content to be reasonably understandable.
Other researchers have observed similar patterns, like Bravo et al.~\cite{bravo2010bridging}, who found that even simplified policies can sometimes fail to achieve universal clarity.
Despite this limitation, our findings are consistent with evidence that structured, plain-language approaches are effective in bridging the comprehension gap~\cite{milne2004consumers, schaub2015design}.

\subsection{Which privacy concepts received low understandability ratings?}
\label{secUnderstandabilityHiLo}

\begin{table}[t]
	\centering
    \footnotesize
	\begin{tabular}{@{}l | l | l@{}}   
    \textbf{Case Label} &
    \textbf{U mean} &
    \textbf{U SD}\\\hline\hline
    
\caseTable{89}{The service has a no refund policy} &
9.67	& 0.82\\\hline

\caseTable{154}{Users who have been permanently banned from this service are not allowed to re-register under a new account}&
9.67	& 0.50\\\hline

\caseTable{175}{You are responsible for any risks, damages, or losses that may incur by downloading materials} &
9.60	& 0.55\\\hline

\caseTable{131}{This service is only available for use individually and non-commercially} &
9.57	& 1.13\\\hline

\caseTable{191}{You can retrieve an archive of your data} &
9.56	& 1.01\\\hline

\caseTable{27}{If you are the target of a copyright claim, your content may be removed}&
9.50	& 0.55\\\hline

\caseTable{218}{Your browsing history can be viewed by the service} &
9.43	& 1.13\\\hline

...& ... &...\\\hline

\caseTable{15}{Content is published under a free license instead of a bilateral one	} &
3.50	& 2.28\\\hline

\caseTable{19}{Defend, indemnify, hold harmless; survives termination} &
3.31    & 2.72\\\hline
\caseTable{80}{The service claims to be CCPA compliant for California users} &
3.00	& 2.83\\\hline
\caseTable{194}{You cannot delete your account of this service} &
2.82	& 2.68\\\hline

\caseTable{81}{The service claims to be GDPR compliant for European users} &
2.67	& 2.78\\\hline

\caseTable{76}{The policy refers to documents that are missing or unfindable} &
2.63	& 0.74\\\hline

\caseTable{155}{very broad} &
1.70	& 1.34\\\hline

	\end{tabular}
	
    \normalsize
	\caption{Descriptive statistcs for the top and bottom ranked cases, based on mean \textit{understandability} (U).
    Please refer to our supplemental materials for descriptive statistics for all cases.
    }
	\label{tableConcUnderstand}
\end{table}

\boldify{Next we zoom into cases with low understandability, which matter because they are opportunities}

Cases with low understandability ratings indicate opportunities to clarify the case or educate the public.
} 
Table~\ref{tableConcUnderstand} shows the text and ratings of the privacy concepts that participants found most (and least) understandable.
Since the overall understandability was high, only about 11\% of the responses had a low understandability score of less than 3.
In terms of number of cases, only 5 cases had an understandability score of 3 or less.
Fortunately, the corresponding \textit{severity} scores of these cases was between -3.3 to 2.72, suggesting these cases are not very serious.
Our results show that a portion of the participants had difficulty understanding certain privacy concepts, despite the fact that the average score of 7.52 and 85.01\% of scores being 5 or higher indicates that the majority of respondents found the content understandable.

\boldify{Let's look at a case where clarification is due: word salad}

We found some aspects of this data unsurprising, such as \case{155}{very broad} obtaining a very low understandability score, suggesting clarification is warranted for this case.
Note that this case label actually implies that users \textit{provide platforms broad rights} to use, alter, and distribute their material without any additional limitation.
This provision curtails users' autonomy in determining the handling of personal data.

\boldify{And one more interesting one: missing documents/broken links}

Contrastingly, we were surprised that \case{76}{The policy refers to documents that are missing or unfindable} received such a low understandability score.
As with the cases found at the top of Table~\ref{tableConcUnderstand}, the terminology is fairly simple, but participants found this case unclear.
Further, they had a strong consensus about this perception, as evidenced by the low standard deviation.
Apparently, this particular case should be fairly common, as recent work found that \textit{``privacy policies URLs are only available in 34\% of websites''}~\cite{srinath2023lost}.

\subsection{Which privacy concepts exhibited low consensus about understandability?}

\begin{table}
	\centering
    \footnotesize
	\begin{tabular}{@{}l | l | l@{}}   
    \textbf{Case Label} &
    \textbf{U mean} &
    \textbf{U SD}\\\hline\hline
    
\caseTable{147}{Tracking pixels are used in service-to-user communication}&
3.75	& 3.92\\\hline

\caseTable{4}{Accessibility to this service is guaranteed at 99\% or more}&
5.40	& 3.91\\\hline

\caseTable{197}{You cannot opt out of promotional communications}&
4.42	& 3.90\\\hline

\caseTable{44}{Minors must have the authorization of their legal guardians to use the service}&
6.25	& 3.86\\\hline

\caseTable{6}{Alternative accounts are not allowed}&
5.67	& 3.82\\\hline

\caseTable{217}{Your browser's Do Not Track (DNT) headers are respected}&
5.56	& 3.81\\\hline

\caseTable{71}{The copyright license that you grant this service is limited to the parties that make up the service's broader platform}&
5.45	& 3.78\\\hline

...& ... &...\\\hline

\caseTable{43}{Many third parties are involved in operating the service}&
7.00	& 0.76\\\hline

\caseTable{25}{First-party cookies are used}&
9.17	& 0.75\\\hline

\caseTable{76}{The policy refers to documents that are missing or unfindable}&
2.63	& 0.74\\\hline

\caseTable{79}{The service can intervene in user disputes}&
6.00	& 0.63\\\hline

\caseTable{175}{You are responsible for any risks, damages, or losses that may incur by downloading materials}&
9.60	& 0.55\\\hline

\caseTable{27}{If you are the target of a copyright claim, your content may be removed}&
9.50	& 0.55\\\hline

\caseTable{154}{Users who have been permanently banned from this service are not allowed to re-register under a new account}&
9.67	& 0.50\\\hline

	\end{tabular}
	
    \normalsize
	\caption{
    Descriptive statistics for the top and bottom ranked cases, based on standard deviation of understandability (U)
    }
	\label{tableDispUnderstand}
\end{table}

\boldify{Now we switch lenses to examine consensus as a way to triangulate with central tendency}

Large standard deviations indicate variation in the participants' ability to understand particular ideas.
This implies that some participants may have found some elements of the privacy material more difficult to understand than others.
Although participants perceived most cases to be moderately understandable, 45 cases had a standard deviation of 3 or more.
Table~\ref{tableDispUnderstand} illustrates the cases that had the most and least consensus about understandability.

\boldify{Variation among participants can explain some of this, but also could indicate opportunities to clarify/educate}

Notably, several of the cases found in the top portion of Table~\ref{tableDispUnderstand} involve technical content (e.g., tracking pixels, do not track headers) or copyright issues.
As a result, we might expect natural variations in the understandability of these cases due to variations in participants' technical and legal knowledge.
On the other hand, the bottom portion of the table tends to refer to more accessible topics (e.g., responsibility, content removal, number of parties involved).

\revised{%
\subsection{How did participants rewrite case labels for low understandability concepts?}

\boldify{Give a little context about the overall data density of the exercise}

Out of 243 unique cases, participants submitted rewrites for 215 cases (88\%).
Out of those 215 unique cases, 72 cases had only 1 rewrite and 63 cases had 2 rewrites.
As the number of rewrites increases, the number of cases decreases, with only 4 unique cases receiving 6 rewrites.
Table~\ref{tableRewwriteStats} illustrates the case labels of the cases that got the most rewrites with the least consensus defined by the standard deviation score of understandability of those cases being 2 or more.
For these cases, participants perceived a larger severity range, as evidenced by standard deviation scores of up to 7.

\boldify{JED is not totally sure, but it seems to me that the idea is that both of these participants generated similar text, which is essentially an expansion of the list}

Participants rephrased this case:
\case{192}{You can request access, correction, and/or deletion of your data} by essentially expanding upon the three-item list provided.
} 
\quotateInset{P146}{ignore}{ignore}{You are allowed to ask the service provider to access your data that it has collected, to correct the data, or to have it deleted from their system altogether. }
\quotateInset{P37}{ignore}{ignore}{You can see your personal data that has been collected. If there are any errors, you are able to make corrections. You can also have your data deleted completely.}

\begin{table}
	\centering
    \footnotesize
	\begin{tabular}{@{}l |l|l|l@{}}   
    \textbf{Case \#} & \textbf{rewrite \#} &
    \textbf{U SD} &
    \textbf{Severity SD}\\\hline\hline

\caseTable{192}{You can request access, correction and/or deletion of your data} & 6 &
3.71	& 2.30	\\\hline

\caseTable{153}{Usernames can be rejected or changed for any reason} & 6	& 3.52	& 5.13
\\\hline

\caseTable{10}{Any liability on behalf of the service is only limited to the fees you paid as a user} & 5	& 3.08	& 3.16
\\\hline

\caseTable{2}{A license is kept on user-generated content even after you close your account} & 5	& 2.92	& 5.36
\\\hline

\caseTable{36}{Instead of asking directly, this Service will assume your consent merely from your usage.} & 5	& 2.77	& 7.01
\\\hline

\caseTable{94}{The service is only available in some countries approved by its government} & 6	& 2.16	& 2.97
\\\hline

\caseTable{125}{This service gives your personal data to third parties involved in its operation} & 5	& 2.12	& 1.67
\\\hline

\caseTable{201}{You have the right to leave this service at any time} & 6	& 2.00	& 3.78
\\\hline

	\end{tabular}
	
    \normalsize
	\caption{ Cases with maximum rewrites having low understandability (U) consensus, alongside  
    }
	\label{tableRewwriteStats}
\end{table}

\revised{%
To take a second example, participants rewrote this case:
\case{153}{Usernames can be rejected or changed for any reason} in a similar way, expanding upon the provided label and removing passive voice.
} 
\quotateInset{P473}{ignore}{ignore}{The company can, without notifying the user in advance, change or delete screen usernames at any point for any reason.}
\quotateInset{P29}{ignore}{ignore}{The service provider can change your username without having to notify you.}

Despite the case label not incorporating the element of notification, both of these participants drew it out of the more detailed description, which suggests that clarification is important to them.

\subsection{Implications}

\boldify{\#1 We can explain with this taxonomy without doing further work}

First, the overall understandability of the cases is fairly high, \textbf{suggesting that \tosdr{} cases are suitable for use as explanation artifacts with little to no modification}.
That said, there is clearly still some room to improve the clarity of the text.
This is evident because ratings were not uniformly high and because we were able to find some rewrites that we thought were better.

\boldify{\#2 By following two sources, we can identify things to spur education and clarification (i.e., we might still want to do some revisions by accepting some of the work)}

Second, we have identified concepts that may be appropriate for educational interventions.
Specifically, the cases with low overall understandability indicate that either the label is confusing or that participants are under-informed about that concept.
Further, the cases with low consensus about the level of understandability indicate that \textit{certain segments} of the sample are under-informed about that concept (we will return to this topic in Section~\ref{secRQ4}).
Last, identifying concepts where expert opinion diverges from participants' perception of the party a concept favors and how much (severity) could prove a valuable guide for educational intervention.
As such, we turn to severity data next.

\section{Results RQ2 - What does the \textit{severity} data reveal about participants and privacy concepts?
\draftStatus{one fix remains, revision highlights are in}
}

\revised{%
Next, we follow a similar approach as in the last section, but turn our focus to the \textit{severity} data.
As mentioned in Section~\ref{secAnalysis}, we will at times refer to severity and other times the absolute value of severity, which we will denote $|S|$.

\boldify{Here we want to make two points: 1. cases favor companies; 2. we see little consensus on severity}

Based on the 2495 responses by participants, they seemed to \textbf{perceive cases as broadly in favor of Service Providers}.
Only one-third of the responses indicated that the cases were in favor of the clients/participants, while over half (53\%) of the responses indicate that the cases favor the Service Providers and about 14\% responses were neutral.
Participants typically evaluated cases as having a moderate level of severity, with a mean severity of -2.23 and a median of -2.00.
However, \textbf{there was little overall consensus in severity}, as indicated by a standard deviation of 6.07.
This aligns with prior research by Bansal et al.~\cite{bansal2010impact}, which highlights that users tend to view privacy policies as skewed in favor of organizations, particularly when transparency is limited or trust is compromised. 


\subsection{Which privacy concepts received high severity ratings?}
\label{sec:highSeverity}

\begin{table}[b]
	\centering
    \footnotesize
	\begin{tabular}{@{}l | l | l@{}}   
    \textbf{Case Label} &
    \textbf{S mean} &
    \textbf{S SD}\\\hline\hline

\caseTable{138}{This service reserves the right to disclose your personal information without notifying you} &
-9.75	& 0.50\\\hline

\caseTable{218}{Your browsing history can be viewed by the service} &
-9.71	& 0.76\\\hline

\caseTable{64}{Service fines users for Terms of Service violations} &
-9.57	& 0.79\\\hline

\caseTable{65}{Some personal data may be kept for business interests or legal obligations} &
-9.22	& 0.83\\\hline

\caseTable{156}{Voice data is collected and shared with third-parties} &
-9.08	& 1.16\\\hline

\caseTable{68}{Terms may be changed any time at their discretion, without notice to you} &
-9.00	& 1.95\\\hline

\caseTable{139}{This service retains rights to your content even after you stop using your account} &
-8.90	& 1.20\\\hline

...& ... &...\\\hline

\caseTable{193}{You can scrape the site, as long as it doesn't impact the server too much} &
5.22	& 3.67\\\hline

\caseTable{238}{Your personal data will not be used for an automated decision-making} &
5.25	& 3.22\\\hline

\caseTable{121}{This service does not collect, use, or share location data} &
5.42	& 3.32\\\hline

\caseTable{186}{You can delete your content from this service} &
5.13	& 3.52\\\hline

\caseTable{203}{You maintain ownership of your content} &
5.82	& 4.39\\\hline

\caseTable{230}{Your personal data is not shared with third parties} &
6.18	& 5.38\\\hline

\caseTable{95}{The service is open-source} &
7.14	& 2.97\\\hline

	\end{tabular}
	
    \normalsize
	\caption{
    Descriptive statistics for the top and bottom ranked cases, based on mean \textit{severity} (S).
    Recall that a positive severity score indicates that participants perceived it to favor the User, while a negative severity score favors the Service Provider.
    }
	\label{tableConcSeverity}
\end{table}

\boldify{Two major points here, the polarizing trend, plus the consensus on severity of terms strongly favoring the service provider}

The severity ratings showed a polarizing trend with about 61\% of the responses having an absolute value of 1, 2, 3, 8, 9, or 10 indicating least severe or most severe.
However, there are 6 cases with $|S| \geq 9$, as found in Table~\ref{tableConcSeverity}. 
For cases that strongly favor the Service Provider, there was strong consensus on severity, as indicated by the low standard deviation.

\boldify{What kinds of cases were in the "terms strongly favoring the service provider" bucket}

The cases that strongly favor the Service Provider feature concepts that one might expect, such as accessing private data, events occurring without notice, and sharing data with third parties.
On the other hand, the cases that strongly favor the User were subject to much less consensus, as exhibited by the relatively high standard deviation.
Important concepts visible here include the ``right to be forgotten'' (\case{186}{You can delete your content from this service}), as well as a variety of prohibitions the Service Provider imposes upon themselves (i.e., not sharing, not collecting, not a decision basis). With a relatively high SD of the cases favoring the User, there is a possibility that some participants would have categorized the case favoring the User based on the keyword `You' or `Your' implying the user-friendly condition.
However, previous studies and current surveys have also shown examples where participants have suffered data breaches and data misuse despite some services having their terms favoring the Users~\cite{thomas2017data}.
} 

\subsection{Which privacy concepts exhibited low consensus about severity?}

\begin{table}
	\centering
    \footnotesize
	\begin{tabular}{@{}l | l | l@{}}   
    \textbf{Case Label} &
    \textbf{S mean} &
    \textbf{S SD}\\\hline\hline
    
\caseTable{155}{very broad} &
0.10	& 0.32\\\hline

\caseTable{169}{You are free to choose the type of copyright license that you want to use over your content} &
0.20	& 0.45\\\hline

\caseTable{138}{This service reserves the right to disclose your personal information without notifying you} &
-9.75	& 0.50\\\hline

\caseTable{218}{Your browsing history can be viewed by the service} &
-9.71	& 0.76\\\hline

\caseTable{64}{Service fines users for Terms of Service violations} &
-9.57	& 0.79\\\hline

\caseTable{65}{Some personal data may be kept for business interests or legal obligations} &
-9.22	& 0.83\\\hline

\caseTable{239}{Your personal information is used for many different purposes} &
-8.00	& 1.00\\\hline

...& ... &...\\\hline

\caseTable{96}{The service is provided 'as is' and to be used at your sole risk} &
-4.44	& 7.00\\\hline

\caseTable{233}{Your personal data is used for limited purposes} &
-1.10	& 7.13\\\hline

\caseTable{93}{The service is not transparent regarding government requests or inquiries that may involve your data} &
-4.54	& 7.17\\\hline

\caseTable{175}{You are responsible for any risks, damages, or losses that may incur by downloading materials} &
-3.00	& 7.21\\\hline

\caseTable{232}{Your personal data is used for advertising} &
-4.25	& 7.23\\\hline

\caseTable{41}{Logs are kept for an undefined period of time} &
-2.83	& 7.41\\\hline

\caseTable{103}{The service will resist legal requests for your information where reasonably possible} &
0.00    & 7.54\\\hline

	\end{tabular}
	
    \normalsize
	\caption{
    Descriptive statistics for the top and bottom ranked cases, based on standard deviation of severity (S)
    }
	\label{tabledispSeverity}
\end{table}

Cases that have a high standard deviation of severity scores indicate a lack of consensus about the impact of that concept.
Recall that the data transformation described in Section~\ref{secAnalysis} will have the effect of increasing standard deviation beyond what we might expect for understandability, due to the increased range.

The absolute values of severity ratings exhibit a mean of 5.91 and a median of 6.00, suggesting that participants often assigned a moderate level of seriousness to privacy concerns on the scale.
However, the standard deviation of 3.04 indicates a substantial amount of variation, indicating that individuals held inconsistent views on the severity of the cases.
Table~\ref{tabledispSeverity} shows the cases with the greatest and least consensus.

Notably, the top half of Table~\ref{tabledispSeverity} shows that cases with strong consensus seemed to have two flavors: those strongly favoring the Service Provider and those favoring neither party.
Interestingly, participants do not seem to be accurately assessing the impact of \case{155}{very broad}.
We have discussed this case before, back in Section~\ref{secUnderstandabilityHiLo}, where we described how it strongly favors the Service Provider.
To some extent, such misperception is unsurprising since that case received the lowest understandability score.
Section~\ref{secRQ3} returns to the relationship between severity and understandability.

Turning to the lower half of Table~\ref{tabledispSeverity}, we see that many of the cases with high standard deviation have means closer to the middle of the [-10, 10] range.
This indicates that although certain participants interpreted a case as being in favor of one party, others regarded it as favoring the \textit{other} party.

\subsection{Implications}

First, the \textbf{severity ratings reflect strong power dynamics inherent in privacy regulations}, as many cases appear to favor the protection of the Service Provider over the User.
The polarized ratings of severity clearly indicate that firms tend to put safeguarding their own interests over ensuring consumer privacy.
This is unsurprising in some ways since the Service Provider is the one creating the document in the first place.

Second, \textbf{many cases had very low consensus for severity ratings}.
Certain participants saw these cases as extremely severe (favoring the Service Provider), while others considered them to be less influential, or even favoring the User.
We attribute the limited agreement on severity to two primary contributing factors.
The first is \textit{contrasting viewpoints on the importance of privacy}.
Various participants have different views on the importance of different categories of data and protective techniques.
Although some individuals may not perceive certain privacy regulations as detrimental, others consider them to be grave infringements.
The second is \textit{deficiencies in complexity and comprehensibility}.
Certain privacy terminology may need specialized expertise to appreciate its consequences.
Should participants have limited technical knowledge, they may struggle to precisely assess the seriousness, resulting in increased inconsistency in evaluations.
Thus, we turn to the interactions between severity and understandability next.

\section{Results RQ3 - How do severity and understandability relate to each other?
\draftStatus{one fix remains, revision highlights are in}
}
\label{secRQ3}

\revised{%
Here, we examine the relationship between severity and understandability from multiple perspectives, including via correlation and analyzing the extremities of the range from our data collection.
For each question, we cropped the data to meet the specific condition for that question.

\begin{figure}
  \centering
  \includegraphics[width=.85\textwidth]{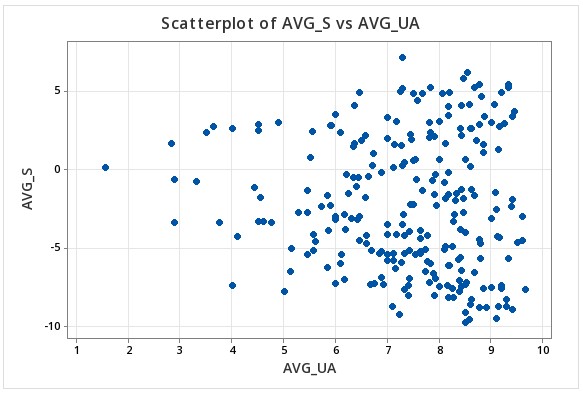}
  \caption{
    Scatterplot of Average Severity (AVG\_S) vs Average Understandability (AVG\_UA).
  }
  \Description{The scatter plot is a reflection of the varied responses for the average values of understandability and severity. As understandability increases, the severity shows greater variability, indicating that participants respond with a wider range of severity for cases they find more understandable.}
  \label{fig:scatterplot-ua-vs-s}
  \FIXED{JED: List of demands for this figure:
  1. export this figure as a pdf, this is a bitmapped representation and is thus bad for zooming
  2. Remove the chart title so that it fits nicer on the page
  3. Relabel things from UA to U, since that is the notation we use elsewhere in the paper.
  4. Clarify in the caption what a dot is here (I think it is a case)
  }
\end{figure}

Analysis of participant ratings about case severity suggests that \textbf{increased understandability scores weakly correlate with elevated mean severity scores}.
Figure~\ref{fig:scatterplot-ua-vs-s} shows the scatter plot of these data, which demonstrate a weak positive correlation (Pearson, 0.154) between participants' perception of understandability and severity.
In addition, the data illustrate a ``funnel shape'', with very few samples with extreme severity ratings paired with low understanding.

\begin{figure}
    \centering
    \includegraphics[width=0.8\linewidth]{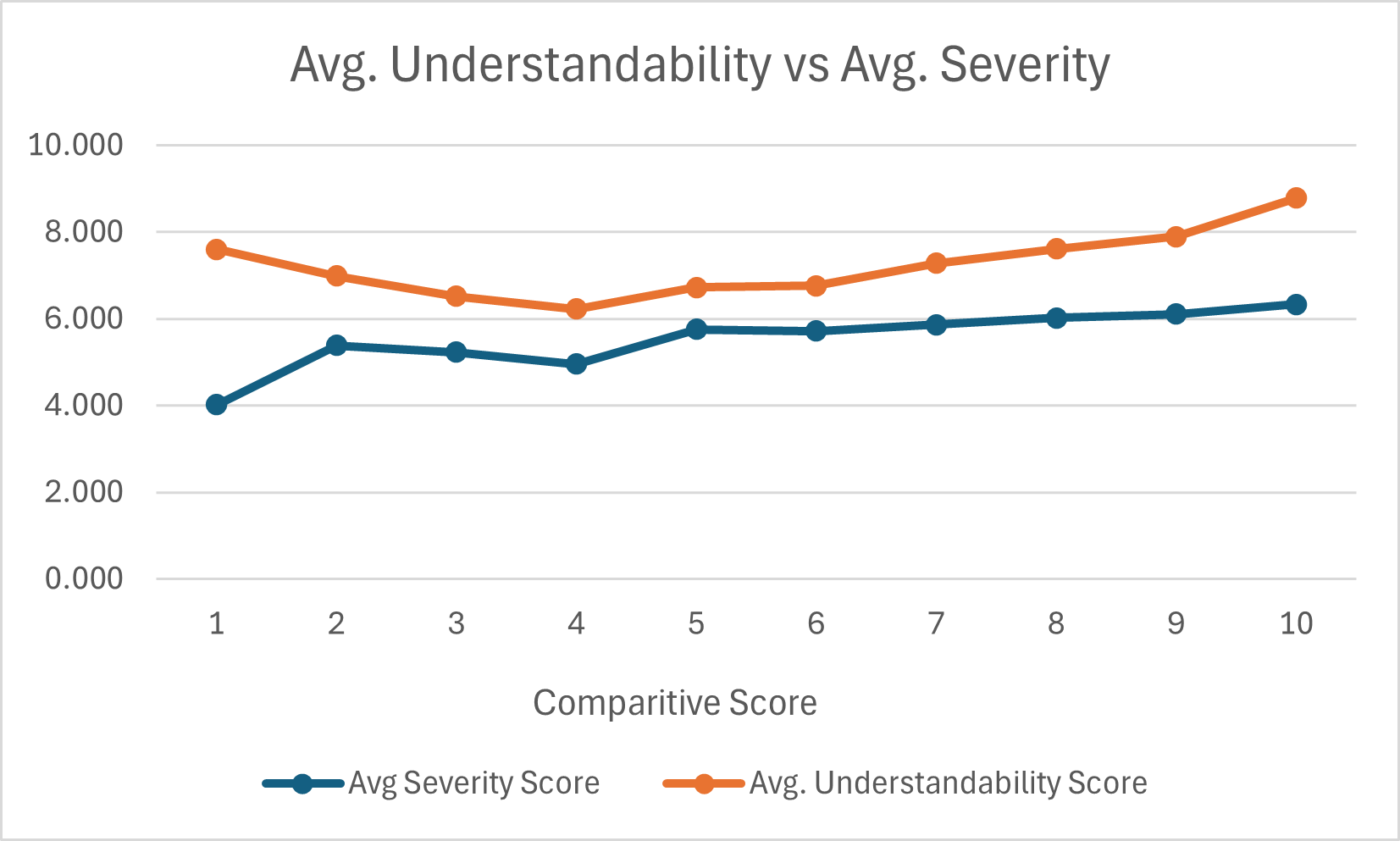}
    \caption{Relationship between average Understandability scores and average Severity scores.
    To read the figure, when the understandability score was 1, the mean severity score was 4, and when the severity score was 1, the mean understandability score was about 8, and so on.
    Note that as the understandability score increases from 1 to 10, the average severity score correspondingly rises from 4 to 6.3.
    }
    \Description{The figure is of a line chart, with two data series, one for understandability, the other severity. Each categorical X-axis position is given by a rating for one of the data series. On the numerical Y-axis is the average value of the OTHER data type for all cases receiving the X-axis rating. In the figure, we broadly see a trend for understandability that starts high, goes down, then back up. Severity starts low and goes up before plateauing.}
    \label{fig:relationship}
    \FIXED{JED@Anyone: List of demands for this figure:
    1. There is a spelling error in the X-axis label (comparative)
    2. Put the legend in the chart area so that we can make the image smaller on the page
    3. specify the Y axis to be from 2 or 3 to 9 or 10 so that we zoom in on the relevant portion a bit more.
    4. Widen the chart area, so it fills the page a bit better by having a wider aspect ratio.
    5. Remove the title

    Finally,
    6. Export this figure as a pdf, this is a bitmapped representation and is thus bad for zooming.
    7. Clarify and extend what I wrote in the caption since I am guessing this is on the $|S|$ data, as opposed to $S$, otherwise the ranges would be different. Without being sure, it is hard for me to finish the last part of the caption (what the reader is supposed to see in the figure)
    }
\end{figure}

Figure~\ref{fig:relationship} shows the trend of average understandability and severity scores with respect to  each other.
As the severity scores go up, the mean understandability scores also go up, a pattern reflecting the observations from prior work.
In one case, Reidenberg et al.~\cite{reidenberg2015disagreeable}, who found that users tend to engage more critically with privacy cases when the content is simplified and the implications are easier to grasp.
Our results also align with Kumar et al.~\cite{kumar2020strengthening}, who observed that when individuals comprehend privacy concepts, they often express concern about the potential harms, particularly when the implications are framed as directly affecting their rights or interests.

In Section~\ref{sec:highSeverity} we noted that severity is polarized towards values of 1 and 10 favoring both parties. 
Here we note that the mean understandability also follows the polarity, with the extreme cases being the most understandable.
Bravo et al.~\cite{bravo2010bridging} reported similar polarization in user reactions to annotated privacy policies, where simplified summaries often elicited strong opinions even with omission of detail.
Our findings indicate that participants are able to better assess severity once the case is more understandable.

\subsection{How severe are the minimally understandable cases?}

In order to determine the ``minimally understandable'' cases, we selected the responses with an understandability score of `1' (110/2495).
When examining the these cases, the data unveils a more intricate structure.

Out of the minimally understandable responses, a total of 35 responses assigned a severity rating of -5 or lower to the case indicating that they considered the cases to be severe.
All of these cases were in favor of the Service Provider. 
This indicates that despite challenges in understanding the information, some individuals still perceived the implications of the case as severe for the user.
Sundar et al.~\cite{sundar2013unlocking} noted similar patterns, where users relied on heuristic decision-making to assess the implications of privacy scenarios, even when comprehension was low.
Further, a total of 7 responses considered the cases to be most severe and favoring the Service Provider.
One such example of a case is 
\case{71}{The copyright license that you grant this service is limited to the parties that make up the service's broader platform.}
These findings align with Isaak and Hanna~\cite{isaak2018user}, who observed that users can intuitively recognize harm in privacy practices, even when they lack full understanding.

\subsection{How severe are the maximally understandable cases?}

We then examined the most understandable instances, in which participants reported `10' understandability (814/2495).
Of these, 424 instances were in favor of the Service Provider, out of which 200 also received the most extreme severity level of -10.
Conversely, only 284 responses (of the 814 maximally understandable cases) indicated a perception of being in favor of the User, with 59 cases classified as the most severe (severity = 10).

Our findings indicate that individuals who have a complete understanding of a case are more likely to gauge it as extremely severe, especially when the Service Provider gains advantages from the case.
This observation aligns with the work of Kelley et al.~\cite{kelley2010standardizing}, which showed that clarity in privacy policies helps users recognize severe consequences tied to specific practices.

\subsection{How understandable are the most extremely severe cases?}

Furthermore, we investigated the understandability of the most extremely severe cases, particularly concentrating on situations when participants assigned the most extreme severity $|S|=10$ (421/2495).
Of the 421 responses participants classified with extreme severity, 344 of those were in favor of the Service Provider, and out of the 344, 200 had the highest understandability score of 10.
These observations suggest that participants well understood a substantial proportion of extremely severe cases.

Notably, 8 responses were simultaneously most extremely severe and minimally understandable.
Among these cases, 7 favored the Service Provider.
These findings suggest that some persons may see a situation as highly serious, even without full understanding, particularly when the outcome is unfavorable to the User.
This reflects Bravo et al.~\cite{bravo2010bridging} and Taddicken \cite{taddicken2014privacy}, who noted that even in cases of limited comprehension, the perceived implications of privacy breaches can significantly influence user perceptions.

} 

\subsection{Implications}

The results obtained from this study have implications for privacy policy formulation and user education.
The substantial association between high understandability and high perceived severity, particularly in cases that benefit the Service Provider, implies that \textbf{users are more inclined to recognize the seriousness of a situation when they possess a stronger understanding of the details}.
However, the existence of severe cases that have low understandability highlights the urgent requirement to improve the understanding of intricate policy terminology, especially those that unfairly advantage Service Providers.
One potential consequence of this lack of understanding is that users may underestimate the potential dangers associated with specific privacy regulations or practices, which in turn highlights the need for focused educational efforts.
Ultimately, more explicit policies not only advantage the user but also establish confidence and promote openness, therefore reducing legal conflicts and enhancing customer well-being.

\section{Results RQ4 - How did demographic traits affect understandability and severity?
\draftStatus{JED about to work here}}
\label{secRQ4}

The survey, conducted through the Prolific platform, got responses from 499 participants\footnote{We had demographic information from the one participant whose survey response was unrecoverable.} from the United States, who came from a very diverse demographic distribution.
Based on the information we collected, we observed the following:

\paragraph{Age} 
Based on the age data, the mean age of the participants is 35.65 years, and half of the participants are below the age of 33.
The standard deviation of 11.44 indicates a considerable heterogeneity in the age distribution among the subjects.
The estimated standard error of 0.51 indicates a little margin of error in calculating the population mean, therefore indicating that the sample offers an accurate estimation of the average age of the participants.
In general, the data indicates a wide range of ages, characterized by a central trend around the mid-30s.

\paragraph{Gender}
The gender distribution in the dataset indicates that women constitute the majority of participants, representing 52.2\% of the total sample. 
Men constitute 46.8\% of the participants, with 1\% opting not to reveal their gender.

\paragraph{Ethnicity}
The slight majority of the participants we sampled identified as White, accounting for 57.6\%.
A total of 13.4\% of the participants identified as Black, 12.2\% identified as Asian, and 10.4\% identified as mixed race.
About 5\% reported belonging to other ethnic groups, while 1\% selected ``N/A'' and 0.4\% opted not to declare their ethnicity.

\paragraph{Employment Status}
A plurality of the participants (36.6\%, 183 individuals) were engaged in full-time employment, while 17.4\% (87 individuals) were working part-time.
Approximately 14.6\% of the respondents reported being unemployed, but an equivalent amount of 14.6\% marked their status as ``N/A'' (which might include students, retired persons, or those who elect not to reveal their employment status).
In addition, 11.2\% of the surveyed population are not currently employed, and 5.6\% are classified as ``Other,'' which may indicate other job circumstances.
The observed distribution exhibits a wide range of job statuses, with full-time workers being the most prevalent category, but also substantial representation from individuals who are not now engaged in paid employment.

\paragraph{Student Status}
The bulk of the participants were not students (67.2\%, 336 individuals), while 19.6\% (98 individuals) self-identified as students.
Furthermore, 13.2\% (66 individuals) replied ``N/A.''

In order to determine how different demographic traits affect understandability and severity, we tested the following  hypotheses:
\begin{itemize}
\item H1: Employed individuals report higher understandability scores than unemployed individuals
\item H2: Older participants record a higher understandability score than young adults.
\item H3: Avg severity scores differ significantly across occupations
\item H4: There is no significant difference in the perceived severity across genders
\end{itemize}

\subsection{H1: Employed individuals report higher understandability scores than unemployed individuals}

In order to test this hypothesis, we categorized participants into two groups:
(1)~\textit{Employed} participants (389 out of 499), who reported their profession or field of work; and
(2)~\textit{Unemployed} participants (110 out of 499), who indicated that they do not work, are students, or selected ``N/A.''
This grouping allowed us to compare understandability scores based on employment status.

We found that \textit{Employed} individuals reported significantly higher understandability scores than \textit{Unemployed} individuals (two-sample t-test, t(161) = -2.93, p = 0.004).
The p-value indicates a statistically significant difference between the groups, rejecting the null hypothesis that employment status has no impact on understandability scores.
The effect size, measured using Cohen's d, was 0.45, which corresponds to a moderate practical significance.
This suggests that while the difference is not only statistically significant, it also has meaningful implications in practical settings.

These findings indicate that employment status plays a role in influencing understandability scores, with \textit{Employed} individuals reporting a better ability to comprehend the material.
This difference may stem from factors such as exposure to workplace communication, problem-solving skills, or familiarity with legal and technical language in professional settings.
Future research could further explore the underlying reasons for this disparity by examining specific aspects of employment, such as job type, education level, or cognitive load, that may contribute to differences in understandability.
Additionally, interventions to enhance understandability for \textit{Unemployed} individuals could help bridge this gap and improve inclusivity in policy and document design.

\subsection{H2: Older participants record a higher understandability score than young adults}

This study hypothesized that older participants would exhibit higher understandability scores compared to younger adults.
Descriptive statistics revealed that participants in the 25--34 age range reported the highest average understandability score (Mean = 7.58, SD = 1.72), while participants aged 65--74 had lower average scores (Mean = 6.17, SD = 2.43).
To test H2, we analyzed understandability scores across different age groups with one-way ANOVA.
We found significant differences among the age groups in average understandability scores (one-way ANOVA, F(6, N=7) = 3.72, p = 0.00125).

These findings suggest that understandability does vary with age, but the trend is not consistent for older participants. These findings suggest that while understandability decreases in older age groups, severity perceptions remain stable across the lifespan.
These results align with prior findings by authors in this article~\cite{lythreatis2022digital}, who noted that younger users often engage more actively with digital platforms but may undervalue privacy risks, while older users tend to exhibit higher privacy concerns due to greater awareness of potential data misuse.
Additional post-hoc tests and a larger sample size should explore this relationship further, especially given the small sample sizes in some age categories (e.g., 75+).

\begin{figure}
    \centering
    \includegraphics[width=0.8\linewidth]{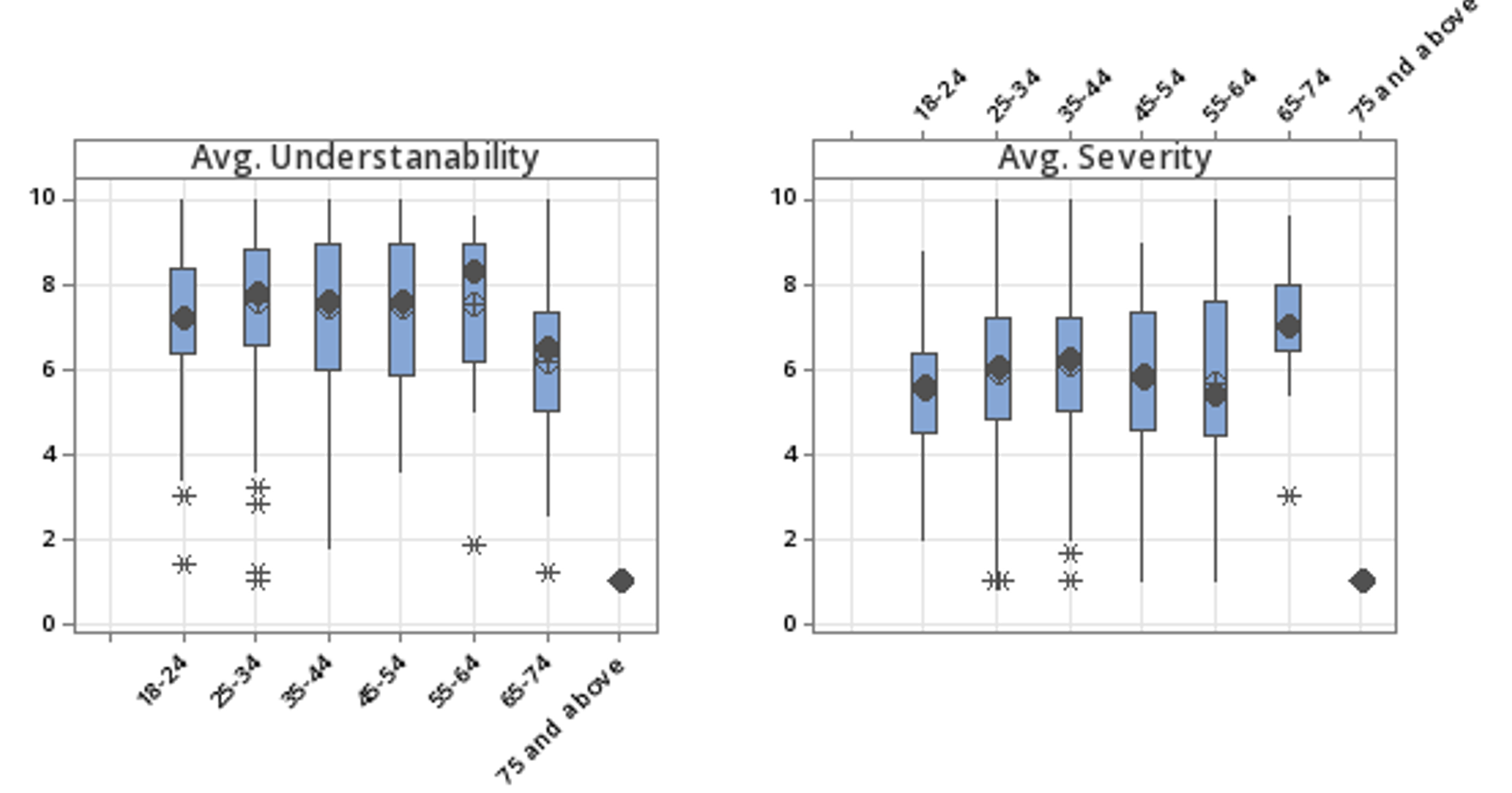}
    \caption{Box Plot for Understandability and Severity scores for different age groups}
    \Description{The figure shows 7 bar charts of each data type broken down by age group. Broadly speaking, the understandability distributions seem to rise with age before peaking and declining. For severity, it is less stable, but generally rises with age.}
    \label{fig:ageBoxplot}
\end{figure}

Figure~\ref{fig:ageBoxplot} highlights the distribution of average understandability and severity scores across age groups.
Understandability scores are relatively consistent among participants aged 18 to 64, with medians around 7--8 and minimal variability, while participants aged 75 and above show a sharp decline in scores.
In contrast, severity scores demonstrate consistent medians (6--7) and narrow interquartile ranges across all age groups, suggesting stability in perceptions of severity.
However, outliers are evident in both metrics, particularly among younger (18--24) and older (75+) participants, indicating individual deviations. 

\subsection{H3: Avg severity scores differ significantly across occupations}

To test this hypothesis, we analyzed average severity scores from participants in various occupational groups via a one-way ANOVA, which examines differences in mean scores across multiple categories.

Descriptive statistics indicated that severity scores showed slight variations among occupations, with mean scores ranging from 5.24 (Arts and Media) to 6.50 (Legal).
Some occupations, such as Legal and Community and Social Services, exhibited relatively high average severity scores with low variability, suggesting more consistent perceptions within these groups.
In contrast, occupations like Arts and Media and Farming, Fishing, and Forestry showed greater variability in scores, indicating a broader range of severity perceptions among participants within these fields.

Despite these observed differences, we found no statistically significant difference in average severity scores across occupational groups (one-way ANOVA, F(29, N=30) = 0.86, p = 0.685).
This finding suggests that occupational background does not have a significant impact on how participants perceive severity.
This can also be a result of a low number of participants leading to a low sample size per occupation.
The results imply that severity perceptions may be influenced more by individual or contextual factors rather than by occupational roles.

\subsection{H4: There is no significant difference in the perceived severity across genders}

To test this hypothesis, we compared average severity scores for all genders via a one-tailed t-test.
Descriptive statistics revealed that men had a slightly higher mean severity score (M = 7.45, SD = 1.87) compared to women (M = 7.37, SD = 1.75).
Participants who preferred not to disclose their gender reported the highest average severity score (M = 8.32, SD = 0.94).
However, the small sample size for the ``Prefer not to state'' was insufficient for statistical testing.

The one-tailed t-test, designed to test whether men report significantly higher severity scores than women, resulted in a t-statistic of 0.49 and a p-value of 0.312.
As the p-value exceeded the conventional threshold for statistical significance (p > 0.05), the analysis failed to provide sufficient evidence to support the hypothesis that men perceive cases as more severe than women.
These findings suggest that, on average, there is no significant difference across these two genders in their perception of severity scores.

While the mean scores for men and women appear similar, the slight variation may be attributed to individual differences, noise in the data, or other confounding variables.
The lack of statistical significance indicates that gender may not be a critical factor influencing the perception of severity in this context.
This result aligns with prior research suggesting that perceptions of severity are often shaped by factors other than gender, such as cultural background, professional experiences, or risk tolerance.
However, these results offer a slight contrast with the prior work by Hoy and Milne~\cite{hoy2010gender}, who observed that women are generally more privacy-conscious and hesitant to share personal information online, while men often prioritize functionality over privacy.

\begin{table}[htb]
    \centering
    \footnotesize
    \begin{tabular}{@{}p{.465\textwidth}|l|l|l|p{.32\textwidth}@{}} 
     \textbf{Questions}
     & \textbf{Mean} 
     & \textbf{Med.}
     & \textbf{SD}
     & \textbf{Interpretation} 
     \\\hline

     \qFour{}
     & 4.08  & 3 & 2.92 &
     Most Participants are not highly motivated to read privacy policies, but there is significant variation, suggesting some individuals are very motivated while others are not.  
     \\\hline

      \qFive{}
      & 4.12	& 3	& 2.88 &
      Participants are generally unlikely to spend time understanding privacy policies, but again, responses varied widely.
     \\\hline
     
     \qSix{}
      & 6.39	& 7	& 3.00 &
      Participants tend to focus more on the general overview rather than specific sections, but the responses vary greatly.
     \\\hline
     
     \qSeven{}
      & 6.63	& 7	& 2.51 & Participants feel moderately confident in managing privacy settings, with relatively less variation in responses.
     \\\hline
     
    \qEight{}
     & 5.87	& 6	& 3.01 &
     Participants feel moderately comfortable using privacy tools, but some are much more comfortable than others.
    \\\hline

    \qNine{}
     & 7.59	& 8	& 2.33 &
     Control over personal information is very important to most participants, with lower variability in responses. 
    \\\hline
    
    \qTen{}
     & 5.39	& 5	& 2.70 &
     Participants are moderately proactive about updating their privacy settings, but responses are somewhat varied.
    \\\hline
    
    \qEleven{}
     & 4.44	& 4	& 2.46 & 
     Participants are generally uncomfortable sharing personal information online, but comfort levels vary across individuals.
    \\\hline
    
    \qTwelve{}
     & 7.06	& 8	& 2.58 &
     Many participants accept terms and conditions without reading them, and this practice is somewhat common across the group.
    \\\hline

    \qThirteen{}
     & 6.84	& 7	& 2.67 &
     Participants are fairly likely to act (e.g., change settings) when concerned about a privacy policy, though responses vary.
\\\hline

     \qFourteen{}
     & 6.48	& 7	& 2.67 &
     Participants show moderate interest in learning about tools to better understand privacy policies, but interest levels vary.

     \\\hline
    \end{tabular}
    \caption{Attitude questions participants saw evaluating privacy concerns with their mean, median and their standard deviation based on the scores each participant provided and the interpretation column to analyze the mean and standard deviation.}
    \label{table: demoStats}
\end{table}

\paragraph{Demographic Questions outlining participant attitudes, based on GenderMag Facets, as applied to privacy attitudes}
Table~\ref{table: demoStats} showcases the responses about the participants' attitudes toward privacy policies and privacy-related behaviors.
It includes 11 questions, each addressing different aspects of privacy management, ranging from motivation to read privacy policies to comfort with sharing personal information.
For each question, the table provides the mean, median, and standard deviation (SD) of responses, along with an interpretation of the results.

The analysis shows that while participants realize the importance of privacy and are somewhat proactive about privacy settings, many are not motivated to read or understand privacy policies.
For example, the mean response for motivation to read policies (Q1) is 4.08, indicating that participants are generally not inclined to invest time in reading, though individual responses vary widely.
This aligns with findings by Acquisti et al.~\cite{acquisti2015privacy}, who showed that the length and complexity of privacy policies discourage users from reading them.
Similarly, participants are moderately confident in their ability to manage privacy settings (Q4) and feel that control over personal information is important (Q6), with relatively lower variability in responses for these items. 
These findings are consistent with Trepte et al.~\cite{trepte2011privacy}, who found that while users value control over their data, their actual knowledge of privacy settings often falls short.

The table also highlights a general discomfort with sharing personal information online (Q8) and a tendency to accept terms and conditions without reading them (Q9), corroborating prior work on user behaviors regarding privacy management.
The authors in this research~\cite{gross2005information} similarly noted that most users are hesitant to share sensitive data but frequently bypass terms and conditions due to convenience.

\begin{figure}[b]
    \centering
    \includegraphics[width=.9\columnwidth]{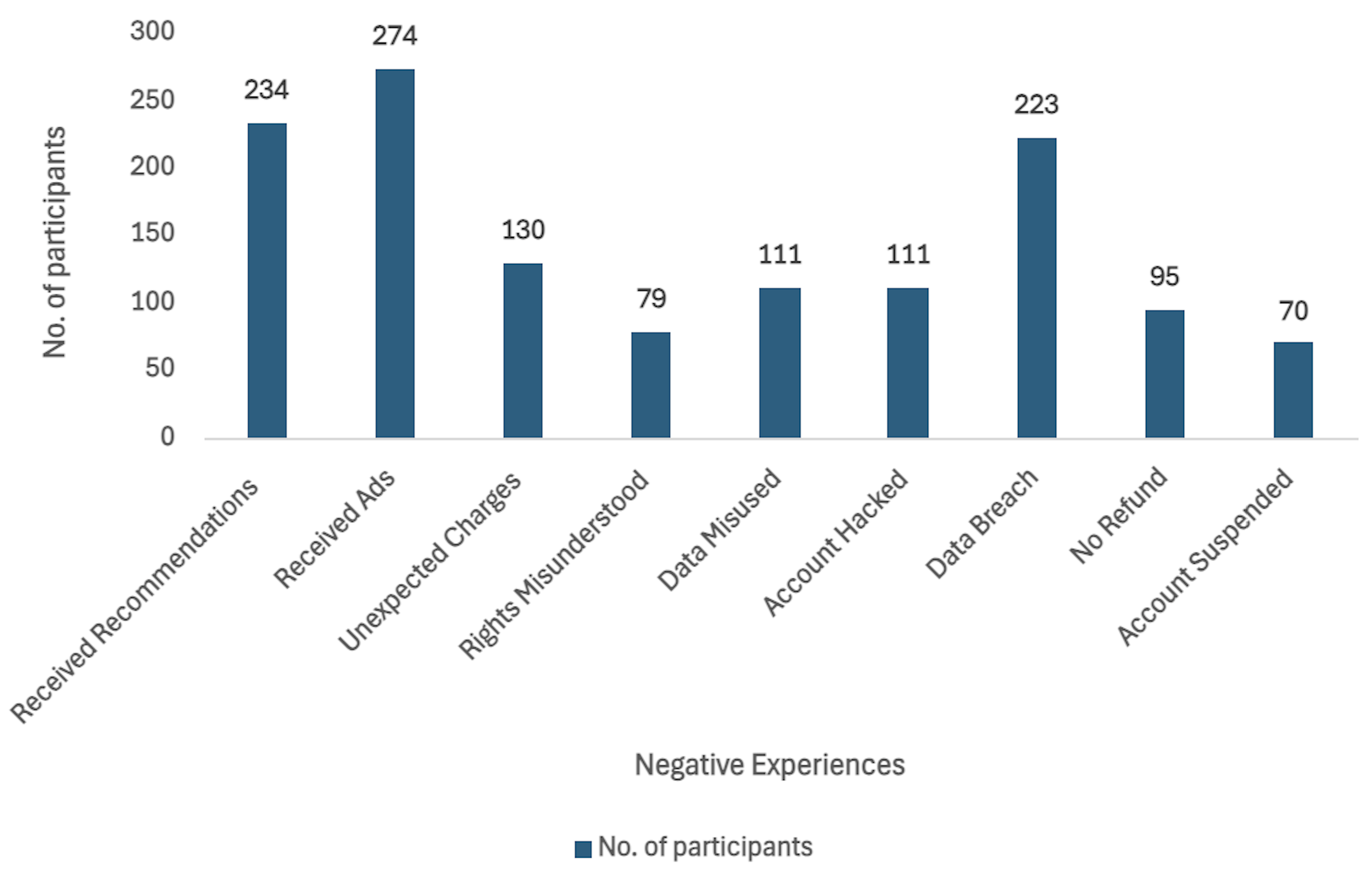}
    
    \caption{No. of participants facing negative experiences}
    \Description{This bar chart shows the trend of how many participants face which negative experiences. As an example, 274 participants have received ads.}
    \label{figNegExp}
\end{figure}

\begin{figure}
    \centering
    \includegraphics[width=.8\columnwidth]
    {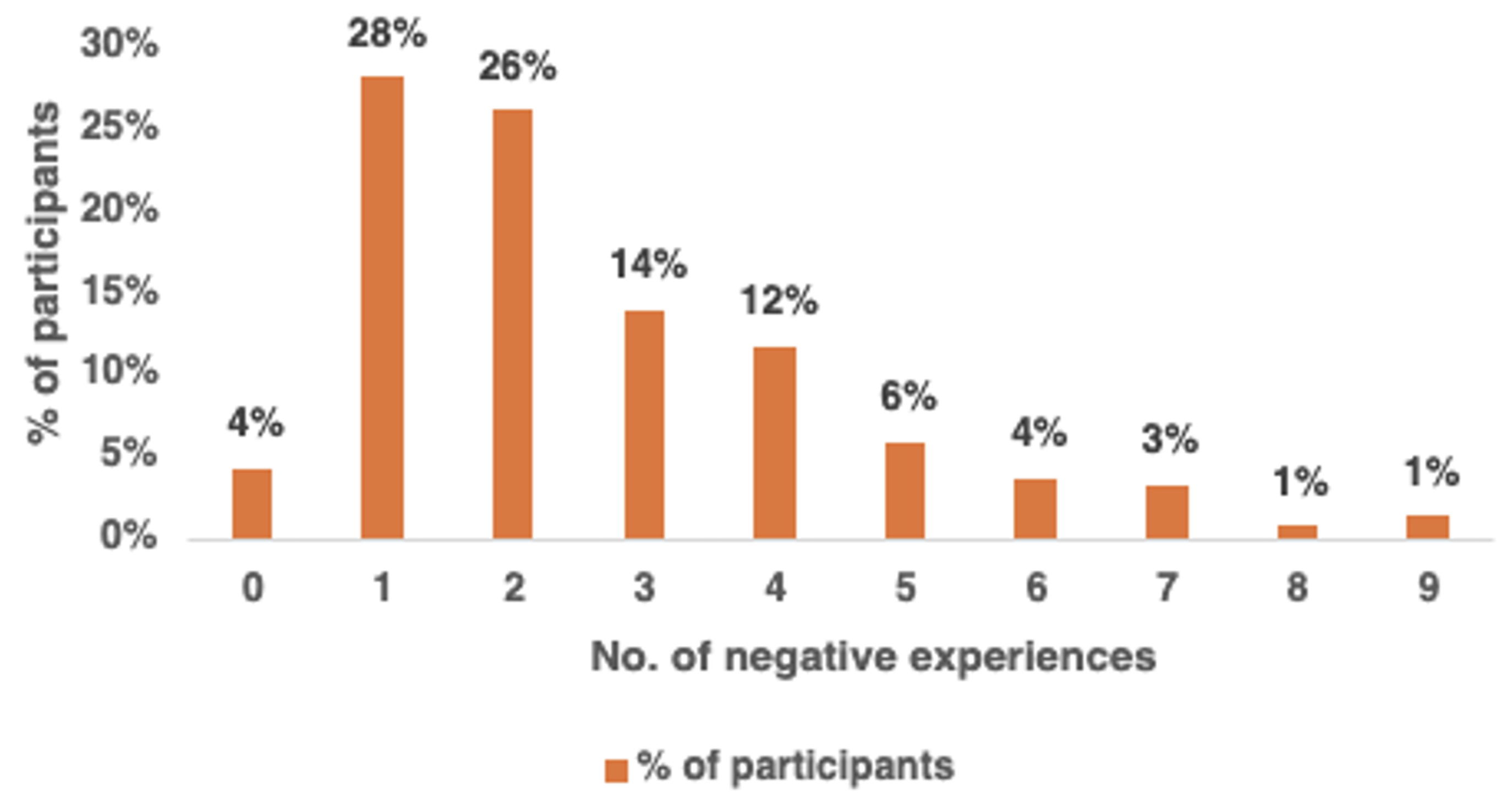}
    
    \caption{Percentage of Participants facing multiple negative experiences}
    \Description{This bar chart shows the trend of how many participants (\%) face a certain number of negative experiences. 4\% of participants have not faced any negative experiences but 28\% faced 1 negative experience.}
    \label{figNegMult}
\end{figure}

\paragraph{Negative experiences}
Figure~\ref{figNegExp} highlights the various categories of adverse experiences encountered by participants, indicating that the most commonly reported concern was the receipt of advertisements, impacting 274 individuals.
This is followed by data breaches reported by 223 participants.
Additional concerns, including unforeseen charges, data exploitation, and account breaches, impacted a considerable number of participants.
Account suspensions were the least prevalent adverse experience, reported by only 70 participants.
These results align with findings by Martin and Murphy~\cite{martin2017role}, who noted that targeted advertisements and data breaches are among the most visible and frustrating outcomes of inadequate data privacy practices.
Figure~\ref{figNegMult} depicts the allocation of participants according to the frequency of negative experiences they faced, indicating that most participants encountered either one or two negative incidents.
The proportion of people indicating three or more adverse events consistently declines, with a mere 1\% reporting eight or nine occurrences.
Notably, 4\% of individuals reported no adverse events.
This aligns with findings by Solove~\cite{solove2012introduction}, who documented that privacy breaches are increasingly common but often concentrated among a subset of highly affected users.


\revised{\section{Discussion
\draftStatus{barring audits, R+R DONE}
}

\subsection{What value does a user-centered weighting for an expert taxonomy provide?}

\boldify{Illuminates participants' values, and so has value in its own right, as in Grgic-Hlaca}

Illuminating participants' abilities, values, preferences, and so on has value in its own right.
To take a single example, Grgić-Hlača et al.~\cite{GrgicHlaca2018fairness} performed a similar crowdworker survey to collect attitudes about the fairness of using certain features in recidivism prediction.
Such information can support a variety of tasks, such as bug triage informed by human data or feature engineering.

\boldify{Explainability, via weight the presence of these concepts to "grade" a whole policy. Currently, \tosdr{} essentially uses 3 weights , if I recall?)}

Another application for our data is in explainability.
In Section~\ref{secIntroduction} we mentioned how \tosdr{} assigns letter grades to services based on user reports of cases presence in policy documents.
That process currently uses 3 weights: (-10, 0, 10), which is part of the inspiration for our two-part question for severity because the current weights are essentially a sign function.
By adding nuance to the weighting schema, it is possible to improve user experience by both increasing information quality and customization.
Our work can help users focus on concepts that other people have found important (e.g., output can be in order by severity ratings).

\subsection{What are some future research hypotheses or educational interventions that this work illuminates?}

\boldify{inform educational interventions (understandability low, severity differs from expert opinion, low consensus about severity).}

Ultimately, each case that satisfies any of three conditions offer research and intervention opportunities:
Condition 1 - the concept has low understandability;
Condition 2 - the concept has low consensus about severity;
Condition 3 - the severity differs from expert opinion.
While we leave evaluating Condition 3 to future work, our data supports targeting educational interventions and research efforts around concepts meeting Conditions 1 and/or 2.
While our full data is available in our supplemental materials, here we offer two examples of interesting questions to pursue further.

\subsubsection{Why is Open Source such a strong indicator of privacy protection?}

As Table~\ref{tableConcSeverity} shows, the case most favoring Users, by a \textit{wide} margin, is:
\case{95}{The service is open-source}.
We found this interesting because having open-source code does not \textit{inherently} prevent careless or malicious behavior within that source.
However, it does indicate that the Service Provider holds certain values, namely transparency, which apparently participants felt were more favorable than even prohibitions on 3rd party sharing or ownership of their content.

\subsubsection{To what extent do people bound or protected by laws like CCPA/GDPR actually know the rules specified therein?}

\boldify{Let's look at a less clear-cut, but important case: legal compliance}

One surprising result from Table~\ref{tableConcUnderstand} is that the cases following the form:
\case{80, 81, 82}{The service claims to be [LAW] compliant for [REGION] users}
received very low understandability scores.
We offer three hypotheses for this observation:
The first hypothesis is that participants do not know what the GDPR/CCPA \textit{is} because \textit{``acronyms constitute a private language''}~\cite{CISEcareer} and they do not speak it.
The second is that they recognize the acronym, but do not know the contents of the law because it is not relevant to them.
Specifically, our participants were US-based, and therefore not bound by GDPR.
Meanwhile, based on rough population counts, we would only expect about 12\% of our participants to be California-based, and so most of our participants are unlikely to be bound by CCPA \textit{either}.
The last hypothesis is that they recognize the acronym, but do not know the law's contents, \textit{despite} its applicability to their lives.}
\section{Threats to Validity
\draftStatus{barring audits, R+R DONE}
}

We follow the framework from Yin~\cite{yinBook} to analyze threats to validity, since every study has threats~\cite{Wohlin-2012}.

\subsection{Construct Validity - did we actually measure what we thought we measured?}

The first threat in this category is that our case text may not accurately capture the concept it is supposed to represent.
To mitigate this threat, we appealed to a known taxonomy that experts at \tosdr{} curated.
Additionally, our task was to provide perception in a context-free manner.
As a result, the participants may not have felt any sense of stakes, which might exist if we were able to access people who were actually handling a policy document for a real purpose, with the intent of deciding whether or not to agree.
\revised{%
Last, it is important to collect additional information about what each case means for the user or service provider and whom does it \textit{actually} favor.
As such, validated answers from subject matter experts who can confirm or deny the party favored and the severity could provide ground truth.
Note that this is a normative approach, while our work follows a descriptive approach---the combination helps triangulate results.
} 

\subsection{External Validity}

Based on our inclusion/exclusion criteria, we focused on English speakers residing in the US, who were working on the crowdsourcing platform Prolific.
As such, our results definitely do not generalize to other samples (i.e., the standard limitations of frequentist statistics).
\revised{%
Further, crowdsourcing introduces sampling biases, such as those Tang et al.~\cite{Tang2022crowdsourcing} revealed.
The most important one for the present study is that Prolific workers are not representative of the general population with respect to security and privacy knowledge.
} 
However, the hope with doing a study like this is that some findings may generalize to samples that are sufficiently similar to ours.
Thus, it is highly unlikely that our observations generalize to other countries or languages, though it may be possible to generalize to more similar samples, such as English speakers residing in the US who are not working on Prolific).

\subsection{Reliability}

The biggest threat to reliability comes from our sampling procedure.
Ultimately, we wanted to gather data on a total of 243 unique cases.
However, having a single person observe all the cases would result in a very long task, so we randomly chose 5 cases for each participant to review.
This sampling strategy meant that not all cases received equal representation among the responses.
Specifically, while it is possible to write the webserver to cap the number of samples for each case, we have to sample before we assess the work quality.
We could semi-automate this based on our attention check question that we included to improve reliability, but we also based decisions to accept or reject work on the apparent quality of the rest of the responses as well.
As an example, some participants got the attention check correct but provided strange answers and a rewrite that included a ChatGPT prompt.

\subsection{Internal Validity}

While our study is descriptive in nature and thus does not have treatments, we did perform comparative statistics based on splitting the sample.
Hence, all the typical concerns associated with doing so remain.
As an example, employment status may be a proxy for a different, more important variable, such as level of education.

\section{Conclusion
\draftStatus{not great, but barring audits, R+R DONE}
}

Our main findings are as follows.
First, the text describing cases found on \tosdr{} is understandable to laypersons overall, though several cases received low understandability and low consensus on understandability.
Meanwhile, prior work has consistently shown people's confusion and inability to understand whole policies.
This suggests that this text can be explanatory for approaches that use such a taxonomy to classify text.
Second, severity scores seemed to indicate that the cases systematically favor the Service Provider.
Third, many cases had low consensus on severity, which could be a reflection of natural variation in attitudes.
Our results suggest that rewriting those cases may help, but educational interventions about such concepts may help \textit{more}.
Fourth, we observe an interplay between severity and understandability ratings that indicate the need for further study.
By enhancing the clarity and accessibility of privacy terms, organizations have the opportunity to cultivate a more knowledgeable user community, resulting in increased trust and user empowerment in managing privacy-related issues.

\bibliographystyle{ACM-Reference-Format}
\bibliography{0-main}

\end{document}